\documentstyle[11pt]{article} \title{A comparison of Noether charge
and Euclidean methods for Computing the Entropy of Stationary Black
Holes} \author{Vivek Iyer and Robert M. Wald \\Enrico Fermi Institute
and Dept. of Physics\\ University of Chicago\\5640 S. Ellis
Ave.\\Chicago, IL 60637} \date{March 24, 1995}

\begin{document}

\maketitle

\begin{abstract} The entropy of stationary black holes has recently
been calculated by a number of different approaches. Here we compare
the Noether charge approach (defined for any diffeomorphism invariant
Lagrangian theory) with various Euclidean methods, specifically, (i)
the microcanonical ensemble approach of Brown and York, (ii) the
closely related approach of Ba\~nados, Teitelboim, and Zanelli which
ultimately expresses black hole entropy in terms of the Hilbert
action surface term, (iii) another formula of Ba\~nados, Teitelboim
and Zanelli (also used by Susskind and Uglum) which views black hole
entropy as conjugate to a conical deficit angle, and (iv) the pair
creation approach of Garfinkle, Giddings, and Strominger. All of
these approaches have a more restrictive domain of applicability than
the Noether charge approach. Specifically, approaches (i) and (ii)
appear to be restricted to a class of theories satisfying certain
properties listed in section 2; approach (iii) appears to require the
Lagrangian density to be linear in the curvature; and approach (iv)
requires the existence of suitable instanton solutions.  However, we
show that within their domains of applicability, all of these
approaches yield results in agreement with the Noether charge
approach.  In the course of our analysis, we generalize the
definition of Brown and York's quasilocal energy to a much more
general class of diffeomorphism invariant, Lagrangian theories of
gravity.  In an appendix, we show that in an arbitrary diffeomorphism
invariant theory of gravity, the ``volume term" in the ``off-shell"
Hamiltonian associated with a time evolution vector field $t^a$
always can be expressed as the spatial integral of $t^a {\cal C}_a$,
where ${\cal C}_a = 0$ are the constraints associated with the
diffeomorphism invariance.

{\bf PACS \#: } 04.20.-q, 0.4.20.Fy, 97.60.Lf \end{abstract}

\newpage

\oddsidemargin=0in \evensidemargin=0in \topmargin=-.5in
\textwidth=6.7in \textheight=9in \hsize=6.27truein

\newpage

\section{Introduction}

In two recent papers \cite{W}, \cite{IW}, the first law of black hole
mechanics was derived for an arbitrary diffeomorphism invariant
Lagrangian theory of gravity. A simple, general expression for the
entropy of a black hole was thereby obtained, namely \begin{equation}
S = 2\pi\int_{{\cal H}} {\bf Q}[t] \label{S} \end{equation} where
${\cal H}$ denotes the bifurcation surface of the horizon, $t^a$ is
the horizon Killing field, normalised to have unit surface gravity,
and ${\bf Q}$ is the Noether charge $(n - 2)$-form (see section (2)
below).  It was shown in \cite{IW} that the entropy so defined is
also given by \begin{equation} S = -2\pi\int_{{\cal H}} E_R^{abcd}
\mbox{{\boldmath $\epsilon$}}_{ab} \mbox{{\boldmath $\epsilon$}}_{cd}
\label{ER} \end{equation} where $\mbox{{\boldmath $\epsilon$}}_{ab}$
is the binormal to ${\cal H}$ and $E_R^{abcd}$ is the functional
derivative of the Lagrangian with respect to the Riemann tensor,
$R_{abcd}$, with the metric and connection held fixed, i.e.,
$E_R^{abcd}$ is the equation of motion for $R_{abcd}$ which would be
obtained from the Lagrangian if $R_{abcd}$ were treated as a field
{\em independent} of the metric.

The relationship between the Noether charge approach to calculating
the entropy of a black hole and a Euclidean approach first given in
\cite{GH} was already analyzed in \cite{W}. However, recently a
variety of other Euclidean approaches to calculating black hole
entropy have been given. These approaches appear to bear little, if
any, resemblance to the Noether charge approach and, as presented,
they usually have had their range of applicability restricted to
general relativity. Although these Euclidean methods agree with the
Noether charge method in yielding the Bekenstein-Hawking formula
$S=A/4$ in the case of general relativity, this fact alone does not
go far towards establishing that they are equivalent to (or even
directly related to) the Noether charge method.

The purpose of this paper is to examine the relationship between the
Noether charge method and (i) the microcanonical ensemble approach of
Brown and York \cite{BY1,BY2} for general relativity, (ii) the
Hilbert action surface term formula of Ba\~nados, Teitelboim, and
Zanelli \cite{T1}, applicable to general relativity and Lovelock
gravity, (iii) the conical deficit angle formula of Ba\~nados,
Teitelboim and Zanelli \cite{T1} and Susskind and Uglum \cite{SU},
applicable to general relativity, and (iv) the pair creation approach
of Garfinkle, Giddings, and Strominger \cite{GGS}, given for a
particular process in general relativity but applicable, in
principle, to an arbitrary theory. One of our main goals is to
generalize and widen the domain of applicability of these approaches
as much as possible so that they can be compared in a meaningful way
with the Noether charge approach.

In sections 2 and 3, we will examine the microcanonical ensemble
approach of Brown and York \cite{BY1,BY2}.  In section 2, we will
encounter little difficulty in generalizing their notion of
quasilocal energy to an arbitrary, diffeomorphism invariant,
Lagrangian theory for which appropriate boundary conditions have been
specified for variations of the action.  However, in order for the
quasilocal energy to have properties suitable for defining a
microcanonical action, we need to restrict consideration to theories
satisfying certain additional properties specified in section 2.  It
is not clear to us how restrictive these additional properties are,
but we explicitly verify at the end of section 2 that they are
satisfied by general relativity. In section 3, we define a
microcanonical action and show that for the class of theories
satisfying these properties the entropy of a black hole as computed
from this microcanonical action always agrees with eq.(\ref{S}).

The starting point of the Euclidean approach given in \cite{T1} is
essentially the same as that of \cite{BY1,BY2}, but a formula (valid
for general relativity and, more generally, Lovelock gravity) is then
presented which expresses black hole entropy as the limit -- as a
suitable $(n-1)$-surface approaches the horizon bifurcation
$(n-2)$-surface, ${\cal H}$ -- of the surface term appearing in the
Hilbert action of the theory.  In section 4, we give a simple
derivation of this formula for the general class of theories
satisfying the properties listed in section 2.

In section 5, we briefly examine the conical deficit angle formula of
Ba\~nados, Teitelboim and Zanelli \cite{T1}, which also has been used
by Susskind and Uglum \cite{SU}. It appears that the range of
validity of this formula for black hole entropy is fundamentally
limited to theories whose Lagrangians are at most linear in the
Riemann curvature. The relationship between this formula and
eq.(\ref{S}) was previously analyzed by Nelson \cite{Nelson}.

In section 6 we analyze the approach of Garfinkle, Giddings, and
Strominger \cite{GGS} for calculating black hole entropy by comparing
the pair creation rate for black holes (in a process mediated by an
instanton) to the corresponding pair creation rate for monopoles.
Although the language of this approach is extremely different from
that of the Noether charge approach, we shall show that in an
arbitrary theory of gravity, this pair creation rate calculation will
always yield a formula for black hole entropy which is equivalent to
eq.(\ref{S}).

Finally, in the Appendix, we examine the structure of the
``off-shell" Hamiltonian, $H$, arising in an arbitrary diffeomorphism
invariant, Lagrangian theory. We show that $H$ always can be written
as the sum of a surface term plus a volume integral of the form
$\int_\Sigma t^a {\cal C}_a$ where $t^a$ denotes the time evolution
vector field and ${\cal C}_a = 0$ are the constraints associated with
the diffeomorphism invariance.

In the following, we shall follow the notation and conventions of
\cite{Wald}. We will use boldface letters to denote differential
forms on spacetime, and shall, in general, suppress their tensor
indices.

\section{Actions, Hamiltonians, and Quasilocal Energy} In this
section we will consider theories arising from a diffeomorphism
covariant Lagrangian on a manifold $M$ with boundary $\partial M$. We
pose the issue of whether an action, $I_X$, exists for variations
satisfying some (arbitrary) specified boundary conditions (denoted
``$X$") on $\partial M$. We will show that this issue is closely
related to the issue of whether a Hamiltonian, $H_X$, exists when the
same boundary conditions $X$ are imposed upon the fields at $\partial
M$. The value of $H_X$ (when it exists) will then be used to define a
notion of quasilocal energy. We will show that for the case of
general relativity, this notion of quasilocal energy agrees with that
of Brown and York \cite{BY1}, so our analysis may be viewed as a
generalization of the definition of Brown and York to a much wider
class of theories.

We begin by reviewing the elements of Lagrangian field theory which
will be needed for our analysis below. (A much more complete
discussion can be found in \cite {IW}.) We consider theories on an
$n$-dimensional spacetime $(M,g_{ab})$ derived from a diffeomorphism
covariant Lagrangian $n$-form ${\bf L}$ with functional dependence
\begin{equation} {\bf L} = {\bf L} \left(g_{ab}, {\bf
\nabla}_{a_1}R_{bcde},..., {\bf \nabla}_{(a_1}...{\bf
\nabla}_{a_m)}R_{bcde}, \psi, {\bf \nabla}_{a_1}\psi, {\bf
\nabla}_{(a_1}...{\bf \nabla}_{a_l)} \psi\right) \label{L}
\end{equation} where $R_{abcd}$ is the curvature of the connection
${\bf \nabla}$ compatible with the metric $g_{ab}$, and $\psi$
denotes any matter field(s). In the following, we shall we will
collectively denote the dynamical fields, $(g_{ab}, \psi)$, by
$\phi$.  The variation of the Lagrangian defines the equations of
motion form, ${\bf E}$, and symplectic potential ($n-1$)-form ${\bf
\Theta}$ (both of which are local in the dynamical fields) via the
relation \begin{equation} \delta {\bf L} = {\bf E} \delta \phi + d
{\bf \Theta} \label{delL} \end{equation} where ${\bf \Theta}$ is a
function of the dynamical fields and their variations,
\begin{equation} {\bf \Theta} = {\bf \Theta} (\phi, \delta \phi).
\end{equation} As discussed in \cite{IW}, ${\bf \Theta}$ may be
chosen to be covariant (that is, have no dependence on any
``background", non-dynamical fields such as a coordinate system), and
we will assume here that a covaraiant choice has been made.  The
symplectic current, $\mbox{{\boldmath $\omega$}}$, is defined by
taking an antisymmetrized variation of ${\bf \Theta}$,
\begin{equation} \mbox{{\boldmath
$\omega$}}(\phi,\delta_1\phi,\delta_2\phi) = \delta_1{\bf
\Theta}(\phi,\delta_2\phi)-\delta_2{\bf \Theta}(\phi,\delta_1\phi).
\label{omega} \end{equation}

For every diffeomorphism generated by a smooth vector field $\xi^a$
on the spacetime, there is an associated Noether current $(n-1)$-form
${\bf J}$ defined by \begin{equation} {\bf J}[\xi] = {\bf
\Theta}(\phi, {\cal L}_\xi \phi) - \xi \cdot {\bf L} \label{J}
\end{equation} where the centered dot denotes contraction of the
vector into the first index of the form. When the field equations
hold, the Noether current is identically closed (that is $d {\bf J}
\equiv 0$) for all smooth vector fields $\xi^a$, and so can be
written (see \cite{onclosed}) in terms of a globally defined Noether
charge $(n-2)$-form ${\bf Q}$ which is covariant and is locally
constructed out of dynamical fields, \begin{equation} {\bf J}[\xi] =
d {\bf Q}[\xi].  \label{JdQ} \end{equation} The Noether charge form
${\bf Q}$ will play a prominent role in our analysis below.

Now consider the case where the spacetime manifold, $M$, is compact,
with boundary $\partial M$. We define the action, $I$, associated
with ${\bf L}$ by \begin{equation}
 I = \int_M {\bf L} \end{equation} Then, using eq.(\ref{delL}), we
obtain, \begin{equation} \delta I =  \int_M \delta {\bf L} = \int_M
{\bf E} \delta\phi + \int_{\partial M} {\bf \Theta}(\phi, \delta\phi)
\end{equation} Thus, we see that, in general, the action $I$ will
{\em not} be extremized (i.e., will not satisfy $\delta I = 0$ for
all field variations $\delta \phi$) by solutions to the equations of
motion. However, suppose that (for reasons related to quantum physics
or otherwise) we seek a modified action, $I_X$, having the property
that the solutions to the equations of motion are true extrema of
$I_X$ for all field variations $\delta \phi$ satisfying certain
specified boundary conditions (which we denote as $``X"$) on
$\partial M$. (An example of such boundary conditions for general
relativity would be to hold the induced metric on $\partial M$
fixed.) Clearly, a sufficient (and, presumably, necessary) condition
for $I_X$ to exist is for there to exist an $(n-2)$-form
$\mbox{\boldmath $\mu$}(\phi, \delta \phi)$ and an $(n-1)$-form ${\bf
B}(\phi)$ defined on $\partial M$ such that the pull-back,
$\overline{{\bf \Theta}}$, of ${\bf \Theta}$ to $\partial M$ is given
by \begin{equation} \overline{{\bf \Theta}}(\phi,
\delta\phi)\mid_{\partial M} = \delta {\bf B}(\phi) + \overline{d
\mbox{\boldmath $\mu$}}(\phi, \delta\phi)\mid_{\partial M}
\label{defB} \end{equation} for all $\delta \phi$ satisfying the
boundary conditions $X$.  Namely, if we find such a ${\bf B}$, we can
define \begin{equation} I_X = \int_M {\bf L} - \int_{\partial M} {\bf
B} \label{Lprime} \end{equation} and immediately obtain the desired
relation \begin{equation} \delta I_X = \int_M {\bf E} \delta\phi
\end{equation} whenever $\delta \phi$ satisfies conditions $X$ on
$\partial M$.

Now, we relax the condition that $M$ be compact but restrict
attention to the case where $\partial M$ is a timelike hypersurface.
We assume, in addition, that we have foliated $M$ by achronal,
spacelike hypersurfaces, $\Sigma_t$ labeled by parameter $t$, which
intersect $\partial M$ orthogonally in compact $(n-2)$-surfaces,
denoted as $C_t$.  (Note that the requirement that $\Sigma_t$ be
orthogonal to $\partial M$ will impose some restrictions on the
allowed variations of the spacetime metric.) We also assume that we
have chosen a time evolution vector field, $t^a$, which satisfies
$t^a \nabla_a t = 1$ on $M$ and is tangent to $\partial M$. We shall
say that a {\em Hamiltonian} conjugate to $t^a$ exists for the
boundary conditions $X$ if we can find a functional, $H_X$, of the
fields and their derivatives on $\Sigma_t$ such that for all
solutions $\phi$ of the field equations and for all field variations
$\delta \phi$ compatible with the conditions $X$ on $\partial M$, we
have \begin{equation} \Omega(\phi,\delta\phi,{\cal L}_t\phi)=\delta
H_X \label{hameq} \end{equation} where \begin{equation}
\Omega(\phi,\delta_1\phi,\delta_2\phi) = \int_{\Sigma_t}
\mbox{{\boldmath $\omega$}}(\phi,\delta_1\phi,\delta_2\phi).
\end{equation} with $\mbox{{\boldmath $\omega$}}$ defined by
eq.(\ref{omega}). Note, however, that $H_X$ will truly be a
Hamiltonian in the usual sense only when a phase space has been
defined which incorporates the chosen boundary conditions $X$, so
that the $X$-conditions hold automatically for all variations,
$\delta \phi$, in the phase space.  (Even when this is done, there
is, of course, no guarantee that the resulting Hamilton's equations
on phase space will have a well posed initial value formulation, or
even admit any solutions at all.) Finally, note that there would be
essentially no change in the above discussion if we were to replace
the conditions, $X$, on a finite boundary $\partial M$ with suitable
asymptotic conditions at ``spatial infinity" for a manifold $M$
without boundary.

A key point to note here is that there is an intimate relation
between the existence of $I_X$ and $H_X$. As shown in \cite{IW}, we
have \begin{equation} \Omega(\phi,\delta\phi,{\cal L}_t\phi) =
\int_{C_t} \delta {\bf Q}[t] - t\cdot{\bf \Theta}(\phi,\delta\phi)
\label{omeg} \end{equation} whenever the equations of motion hold for
$\phi$ and the linearized equations of motion hold for $\delta\phi$.
Now, suppose that the forms $\mbox{\boldmath $\mu$}$ and ${\bf B}$
can be found so that eq.(\ref{defB}) holds, thereby guaranteeing the
existence of an action $I_X$. Suppose, in addition, that
$\mbox{\boldmath $\mu$}$ satisfies the additional condition that for
all $t$, \begin{equation} \overline{\mbox{\boldmath $\mu$}}\mid_{C_t}
= 0 \label{mu} \end{equation} Then $H_X$ exists and is given by
\begin{equation} H_X = \int_{C_t} {\bf Q}[t] - t\cdot{\bf B}
\label{H} \end{equation} since, when pulled back to $C_t$, we have
\begin{eqnarray} t\cdot \delta {\bf B} & = & t\cdot {\bf \Theta} -
t\cdot d\mbox{\boldmath $\mu$}\nonumber\\ & = & t\cdot {\bf \Theta} -
{\cal L}_t \mbox{\boldmath $\mu$} + d (t\cdot \mbox{\boldmath
$\mu$})\nonumber\\ & = & t\cdot {\bf \Theta} + d (t\cdot
\mbox{\boldmath $\mu$}) \label{thingy2} \end{eqnarray} where
eq.(\ref{mu}) was used in the last step and we omitted writing bars
over all of the differential forms. Eq.(\ref{hameq}) then follows
immediately from eqs.(\ref{omeg}) and (\ref{thingy2}).  Note that
$H_X$ has the surface integral form (\ref{H}) only when the equations
of motion hold for $\phi$.  In the appendix we analyze the ``off
shell" structure of $H_X$.

Suppose, now, that $\mbox{\boldmath $\mu$}$ and ${\bf B}$ can be
found so that eqs.(\ref{defB}) and (\ref{mu}) hold. Equation
(\ref{H}) then motivates the following definitions: We define the
{\em quasilocal energy} (associated with the boundary conditions $X$)
of the cut $C_t$ by \begin{equation} {\cal E}_t \equiv \int_{C_t}
{\bf Q}[u] - u\cdot{\bf B}.  \label{qlcanE} \end{equation} where
$u^a$ is the unit normal to $\Sigma_t$. Similarly we define the
quasilocal momentum conjugate to a vector field $N^a$ tangent to the
cut $C_t$ by \begin{equation} {\cal J}_t \equiv -\int_{C_t} ({\bf
Q}[N^a] - N\cdot{\bf B}).  \label{qlcanJ} \end{equation} As we shall
see below, these definitions generalize the those given by Brown and
York \cite{BY1} for general relativity (see also \cite{Nest}).  Note
also that in the asymptotically flat case, the definitions
(\ref{qlcanE}) and (\ref{qlcanJ}) (for $N^a$ chosen to be an
asymptotic rotation) correspond to the definitions of total energy
and angular momentum given in \cite{IW}.

Although the formula (\ref{qlcanE}) provides a local expression for
an energy-like quantity defined on a 2-surface $C_t$, it should be
noted that this expression would suffer from the following
deficiencies if one were to interpret it as defining a notion of the
energy contained in the region bounded by $C_t$: (1) The definition
of ${\cal E}_t$ depends upon the choice of boundary conditions $X$.
(2) The choice of ${\bf B}$ clearly is ambiguous up to addition of
terms which are constant under field variations which keep $X$
fixed.  (3) The Noether charge ${\bf Q}$ has the ambiguities
discussed in \cite{IW}.  (4) ${\cal E}_t$ does not depend on the
choice of $C_t$ in a suitably continuous manner; specifically, ${\cal
E}$ can take very different values on two surfaces which are very
close to each other in spacetime, but one of which is much
``wigglier" than the other. Note, however, that these difficulties
(1)-(4) do not occur (or, at least, are greatly alleviated) when
defining the total energy or momentum of an asymptotically flat
spacetime.

Until this point, we have considered an essentially arbitrary
Lagrangian theory, with arbitrary boundary conditions $X$ such that
$\mbox{\boldmath $\mu$}$ and ${\bf B}$ can be found so that
eqs.(\ref{defB}) and (\ref{mu}) hold. We now shall restrict
consideration to theories where the following $3$ additional
conditions hold. We shall explicitly verify below that general
relativity satisfies these conditions, but we have not investigated
the precise range of theories for which these conditions (or suitable
generalisations of them) are valid, nor have we even investigated the
extent to which these conditions are independent of each other. For
simplicity, we restrict attention here to the case where matter
fields are absent, so that the only dynamical field is the spacetime
metric $g_{ab}$ \cite{com1}.  \begin{enumerate} \item We assume that
the boundary conditions, $X$, correspond to the fixing of the induced
metric $\gamma_{ab}$ on $\partial M$.  Now, for a fixed choice of
slicing, $\Sigma_t$, and time evolution vector field $t^a$, we can
express $\gamma_{ab}$ in terms of the induced $2$-metric
$\sigma_{ab}$ on the cross-sections $C_t$ together with the lapse,
$N$, and shift $N^a$, defined via \begin{equation} t^a = N u^a +
N^a.  \label{lapseshift} \end{equation} Since $\overline{{\bf
\Theta}} = \delta {\bf B} + \overline{d \mbox{\boldmath $\mu$}}$ on
$\partial M$ when the boundary conditions $X$ are satisfied (see
eq.(\ref{defB})), it follows that for arbitrary metric variations,
$\overline{{\bf \Theta}} - \delta {\bf B} - \overline{d
\mbox{\boldmath $\mu$}}$ at any point $p \in \partial M$ can depend
upon $\delta g_{ab}$ only via $\delta N$, $\delta N^a$, $\delta
\sigma_{ab}$ and their derivatives tangential to $\partial M$. The
freedom available in the choice of $\mbox{\boldmath $\mu$}$ permits
us to eliminate any dependence upon the tangential derivatives, i.e.,
we may always choose $\mbox{\boldmath $\mu$}$ so that on $\partial M$
\begin{equation} \overline{{\bf \Theta}} - \delta {\bf B} =
\mbox{{\boldmath $\alpha$}} \delta N + \mbox{{\boldmath $\beta$}}_a
\delta N^a + \mbox{{\boldmath $\lambda$}}^{ab} \delta \sigma_{ab} +
\overline{d \mbox{\boldmath $\mu$}} \label{thdB} \end{equation} We
now assume that a choice of $\mbox{\boldmath $\mu$}$ can be made
which is simultaneously compatible with both eqs.(\ref{mu}) and
(\ref{thdB}).  \item Consider a slice $\Sigma_t$ and a vector field
$v^a$ on $M$ which vanishes on $\Sigma_t$, so that the
diffeomorphisms generated by $v^a$ leave each point of $\Sigma_t$
invariant. The infinitesimal change, ${\cal L}_v g_{ab}$, induced by
$v^a$ on $g_{ab}$ will, in general, be nonzero on $\Sigma_t$, since
the ``time components" of the metric can change. However, we assume
that these induced variations are ``dynamically trivial" in the sense
that they are degeneracy directions of $\Omega$, i.e., we assume that
\begin{equation} \Omega(g_{ab}, {\cal L}_v g_{ab}, \delta g_{ab})= 0
\label{deg} \end{equation} for all $v^a$ which vanish on $\Sigma_t$
and all metric variations $\delta g_{ab}$.  \item The integrand
appearing on the right side of eq.(\ref{H}) is linear in $t^a$ and
its derivatives. The freedom involved in the choice of ${\bf Q}$
ensures that we may assume that this integrand depends only upon
$t^a$ and its first antisymmetrized derivative \cite{IW}. We now
assume that (using integration by parts if necessary) we can
eliminate the dependence of the integral on the derivatives of $t^a$,
so that we may write eq.(\ref{H}) in the form \begin{equation}
 H_X[t^a] = \int_{C_t} t^a {\bf e}_a \label{Hint} \end{equation}
where ${\bf e}_a$ is independent of $t^a$ and any other background
structure not invariant under diffeomorphisms which leave each point
of $\Sigma_t$ fixed in following sense:  For all metric variations of
the form $\delta g_{ab} = {\cal L}_v g_{ab}$ with $v^a = 0$ on
$\Sigma_t$, we have $\delta {\bf e}_a = {\cal L}_v {\bf e}_a$.
\end{enumerate}

Using the above assumptions, we may relate the quasilocal energy and
momentum to the coefficients $\mbox{{\boldmath $\alpha$}}$ and
$\mbox{{\boldmath $\beta$}}_a$ appearing in eq.(\ref{thdB}). Let
$\hat{\delta}$ denote the variation induced in any quantity by the
variation ${\cal L}_v g_{ab}$ of the spacetime metric, where $v^a$
vanishes on $\Sigma_t$. Then, by eqs.(\ref{omeg}) and (\ref{deg}), we
have \begin{eqnarray} 0 &=& \Omega(g_{ab}, {\cal L}_v g_{ab}, {\cal
L}_t g_{ab}) \nonumber\\ & = & \int_{C_t} \hat{\delta} {\bf Q}[t] -
t\cdot{\bf \Theta}(g_{ab}, {\cal L}_v g_{ab})\nonumber\\ & = &
\hat{\delta} H_X[t^a] - \int_{C_t} (t\cdot{\bf \Theta} -
t\cdot\hat{\delta}{\bf B}) \label{thingy} \end{eqnarray} On the other
hand, using property (3) and $\hat{\delta} t^a=0$, we have
\begin{eqnarray} \hat{\delta} H_X[t^a] & = & \int_{C_t} t^a
\hat{\delta}{\bf e}_a \nonumber \\
  & = & \int_{C_t} t^a {\cal L}_v{\bf e}_a \nonumber \\ & = &
  \int_{C_t} ({\cal L}_v (t^a {\bf e}_a) - {\bf e}_a {\cal L}_v t^a)
  \nonumber \\ & = & - \int_{C_t} {\bf e}_a {\cal L}_v t^a
\end{eqnarray} Finally, using eq.(\ref{thdB}), we obtain from
(\ref{thingy}) \begin{equation} 0 = - \int_{C_t} {\bf e}_a {\cal L}_v
t^a + t\cdot\mbox{{\boldmath $\alpha$}} \hat{\delta} N +
t\cdot\mbox{{\boldmath $\beta$}}_a \hat{\delta} N^a \label{Ltab}
\end{equation} for all $v^a$ which vanish on $\Sigma_t$. Now, using
eq.(\ref{lapseshift}) and the fact that ${\cal L}_v N = {\cal L}_v
N^a = 0$, we have \begin{equation} {\cal L}_v t^a = N {\cal L}_v u^a
= N \hat{\delta} u^a \label{Lievta} \end{equation} On the other hand,
we have \begin{equation} 0=\hat{\delta} t^a = (\hat{\delta}N)u^a +
N\hat{\delta}u^a + \hat{\delta}N^a.  \end{equation} and, consequently
\begin{equation} {\cal L}_v t^a =  -( \hat{\delta}N u^a +
\hat{\delta}N^a) .  \end{equation} Thus, (\ref{Ltab}) implies
\begin{eqnarray} t\cdot\mbox{{\boldmath $\alpha$}} &=& u^a{\bf e}_a,
\nonumber\\ t\cdot\mbox{{\boldmath $\beta$}}_a &=& {\sigma_a}^b {\bf
e}_b.  \label{albe} \end{eqnarray} Thus, we see that, under our above
assumptions, the quasilocal energy and momentum densities are simply
the coefficients of $\delta N$ and $\delta N^a$ appearing in the
pullback to $C_t$ of $t\cdot({\bf \Theta} - \delta {\bf B})$, and we
have \begin{equation} H_X[t^a] = \int_{C_t} Nt\cdot\mbox{{\boldmath
$\alpha$}} + N^at\cdot\mbox{{\boldmath $\beta$}}_a.  \label{Hexp}
\end{equation} This formula corresponds to the definition of the
quasilocal energy and momentum densities given by Brown and York
\cite{BY1}.

We conclude this section by verifying explicitly that all of the
above conditions hold for general relativity. We start with the
expression for the pullback of ${\bf \Theta}$ to $\partial M$ given
in \cite{BW}; \begin{equation} \overline{{\bf
\Theta}}_{abc}|_{\partial M} =
-\frac{1}{16\pi}(K^{mn}-\gamma^{mn}K)\delta\gamma_{mn}
\mbox{{\boldmath $\epsilon$}}_{abc} -\delta(\frac{1}{8\pi}
K\mbox{{\boldmath $\epsilon$}}_{abc}) + \frac{1}{16\pi}d(n^m\delta
n^n \mbox{{\boldmath $\epsilon$}}_{abmn}) \label{thbar}
\end{equation} where $n^a$ is the unit ``outward pointing" normal to
$\partial M$, $\mbox{{\boldmath $\epsilon$}}_{abc} \equiv n^d
\mbox{{\boldmath $\epsilon$}}_{dabc}$ is the induced volume element
on $\partial M$, $K_{ab}$ denotes the extrinsic curvature of
$\partial M$ and $K = {K^a}_a$. It follows immediately that for the
boundary conditions, $X$, of condition (1), the choices
\begin{equation} {\bf B}_{abc} = -\frac{1}{8\pi}\mbox{{\boldmath
$\epsilon$}}_{abc} \left(K + S_0 \right) \label{grB} \end{equation}
and \begin{equation} \mbox{\boldmath $\mu$}_{ab} = \frac{1}{16\pi}
n^m\delta n^n \mbox{{\boldmath $\epsilon$}}_{abmn} \end{equation}
satisfy eq.(\ref{defB}), where $S_0$ is {\em any} quantity only
depending on the intrinsic geometry of $\partial M$.  Furthermore,
using the fact that both $n^m$ and $\delta n^n$ are tangent to
$\Sigma_t$, we see that eq.(\ref{mu}) holds. Since no derivatives of
$\delta \gamma_{ab}$ appear in the first term in eq.(\ref{thbar}), it
also is manifest that eq.(\ref{thdB}) holds, so condition (1) is
indeed satisfied.

Using the form for ${\bf \Theta}$ given in (\ref{thbar}) but now
pulled back to a spacelike slice $\Sigma_t$, one finds
\begin{equation} \Omega(g_{ab},\delta g_{ab},{\cal L}_v g_{ab})  =
\int_{\Sigma_t} \delta\pi^{ab}{\cal L}_v h_{ab} -{\cal L}_v
\pi^{ab}\delta h_{ab} = 0 \label{OmLv} \end{equation} for all $v^a$
which vanish on $\Sigma_t$, where  $h_{ab}$ is the induced spatial
metic on $\Sigma_t$ and $\pi^{ab}$ is its canonically conjugate
momentum. Thus, condition (2) is satisfied.

Finally we show that condition (3) is satisfied in general
relativity.  We have \cite{IW} \begin{equation} {\bf Q}[t]_{ab}=
-\frac{1}{16\pi}\mbox{{\boldmath $\epsilon$}}_{abcd}{\bf \nabla}^ct^d
\label{grQ} \end{equation} and hence \begin{eqnarray} \int_{C_t}
{\bf Q}[t] - t\cdot{\bf B} &=& \int_{C_t}
-\frac{1}{16\pi}\mbox{{\boldmath $\epsilon$}}_{abcd}{\bf \nabla}^ct^d
+ \frac{1}{8\pi}t^c\mbox{{\boldmath $\epsilon$}}_{cab} (K +
S_0)\nonumber\\ &=& \int_{C_t}\mbox{{\boldmath
$\epsilon$}}_{ab}\left(\frac{1}{16\pi}(u_cn_d-u_dn_c){\bf \nabla}^c
t^d +t^au_a\frac{1}{8\pi}(K + S_0) \right) \nonumber \\ &=&
\int_{C_t}\mbox{{\boldmath
$\epsilon$}}_{ab}\frac{1}{8\pi}\left(u_cn_d{\bf \nabla}^c t^d
+t^au_a(K + S_0) \right)\nonumber \\ &&-\int_{C_t}\mbox{{\boldmath
$\epsilon$}}_{ab}\left(\frac{1}{16\pi} (u^cn^d+u^dn^c){\bf \nabla}_c
t_d \right).  \label{sameqle} \end{eqnarray} The last term can be
seen to vanish as follows:  \begin{eqnarray} (u^cn^d+u^dn^c){\bf
\nabla}_c t_d&=&  u^c n^d {\cal L}_t g_{cd}\nonumber\\ &=& u^c {\cal
L}_t n_c - u_c {\cal L}_t n^c \nonumber\\ & = & u^c \left(t^a 2{\bf
\nabla}_{[a}  n_{c]} - {\bf \nabla}_c (t^a n_a)\right)-u_c {\cal L}_t
n^c \label{nodiff} \end{eqnarray} However, the last term on the right
side vanishes on $\partial M$ since $n^c$ is tangent to each
$\Sigma_t$ and $t^a$ generates diffeomorphisms which map the
${\Sigma_t}'s$ into themselves, so ${\cal L}_t n^c$ also is tangent
to each $\Sigma_t$. The middle term vanishes on $\partial M$ since
$t^a n_a = 0$. Finally, using the hypersurface orthogonality of $n_c$
on $\partial M$, we see that the first term also vanishes. Thus, we
obtain \begin{eqnarray} \int_{C_t}{\bf Q}[t]-t\cdot{\bf B}&=&
\int_{C_t}\mbox{{\boldmath
$\epsilon$}}_{ab}\frac{1}{8\pi}\left(u_cn_d{\bf \nabla}^ct^d
+t^du_d(K + S_0) \right)\nonumber \\ &=&\int_{C_t}\mbox{{\boldmath
$\epsilon$}}_{ab}\frac{1}{8\pi}\left(-u^ct^d{\bf \nabla}_cn_d
+t^du_d(K + S_0)\right)\nonumber\\ &=&\int_{C_t}\mbox{{\boldmath
$\epsilon$}}_{ab}\frac{1}{8\pi}\left(-u^ct^d K_{cd} +t^du_d(K +
S_0)\right) \label{finalqle} \end{eqnarray} which is seen to satisfy
condition (3) with \begin{equation} {\bf e}_a = {}^2\mbox{{\boldmath
$\epsilon$}}\frac{1}{8\pi}\left(-u^cK_{ca} +u_a(K +
S_0)\right)\nonumber\\ \label{qldens} \end{equation} We have
therefore verified that general relativity satisfies the conditions
(1)-(3) posed above.

\section{The Microcanonical Action and Black Hole Entropy} Consider a
diffeomorphism invariant theory of a metric $g_{ab}$ derived from an
action, $I_X$, for boundary conditions, $X$, which satisfies the
conditions stated in the previous section. In particular, in such a
theory, the quasilocal energy and momentum densities are defined on
each cut $C_t$.  Following Brown and York \cite{BY2}, we say that an
action $I_m$ is a {\em microcanonical action} for the theory if, for
arbitrary metric variations about an arbitrary metric (subject only
to the restriction that the hypersurfaces $\Sigma_t$ are orthogonal
to $\partial M$), we have \begin{eqnarray} \delta I_m&=&\int_M {\bf
E}^{ab} \delta g_{ab} - \int_{\partial M}dt\wedge
\left(N\delta(u^a{\bf e}_a) +N^a\delta ({\sigma_a}^c {\bf e}_c)
\right) - \mbox{{\boldmath $\lambda$}}^{ab} \delta
\sigma_{ab}\nonumber\\ &=& \int_M {\bf E}^{ab} \delta g_{ab} -
\int_{\partial M} N\delta\mbox{{\boldmath $\alpha$}}
+N^a\delta\mbox{{\boldmath $\beta$}}_a - \mbox{{\boldmath
$\lambda$}}^{ab} \delta \sigma_{ab} \label{deIMBY} \end{eqnarray}
where ${\bf E}^{ab} = 0$ are the equations of motion for $g_{ab}$ and
eq.(\ref{albe}) was used in the second line.  Here, when comparing
signs in our formulas with those of Brown and York, it should be
noted that the relationship between our choice of orientations
${}^{(n-1)}\mbox{{\boldmath $\epsilon$}}$ of $\partial M$ and
${}^{(n-2)}\mbox{{\boldmath $\epsilon$}}$ of $C_t$ is given by
${}^{(n-1)}\mbox{{\boldmath $\epsilon$}} =
-dt\wedge{}^{(n-2)}\mbox{{\boldmath $\epsilon$}}$. (This arises
because we choose ${}^{(n-1)}\mbox{{\boldmath
$\epsilon$}}_{a_1...a_{n-1}} = n^b\mbox{{\boldmath
$\epsilon$}}_{ba_1...a_{n-1}}$ and ${}^{(n-2)}\mbox{{\boldmath
$\epsilon$}}_{a_1...a_{n-2}} = n^bu^c\mbox{{\boldmath
$\epsilon$}}_{cba_1...a_{n-1}}$; the orientation of $\Sigma_t$ is
chosen to be $u^b\mbox{{\boldmath $\epsilon$}}_{ba_1...a_{n-1}}$.

We now show that for theories satisfying the conditions of the
previous section, $I_m$ is given by \begin{equation} I_m = \int_{M}
{\bf L} - \int_{\partial M} dt \wedge {\bf Q}[t] \label{genIM}
\end{equation} i.e., to get $I_m$ we replace ${\bf B}$ in
eq.(\ref{Lprime}) by $dt \wedge {\bf Q}[t]$. Namely, taking the
variation of eq.(\ref{genIM}), we obtain \begin{equation} \delta I_m
= \int_M {\bf E}^{ab} \delta g_{ab}- \int_{\partial M}dt\wedge\left(
\delta {\bf Q}[t] - t\cdot{\bf \Theta}\right) \end{equation} However,
we have \begin{eqnarray} \int_{C_t} \delta {\bf Q}[t] - t\cdot{\bf
\Theta} & = & \int_{C_t}\delta({\bf Q}[t]-t\cdot{\bf B}) -
t\cdot({\bf \Theta} - \delta{\bf B})\nonumber\\ & = & \int_{C_t}
\delta H_X[t] - t\cdot({\bf \Theta} - \delta{\bf B})\nonumber\\ &=&
\int_{C_t} t\cdot\left(\delta (N \mbox{{\boldmath $\alpha$}} + N^a
\mbox{{\boldmath $\beta$}}_a) - \mbox{{\boldmath $\alpha$}} \delta N
- \mbox{{\boldmath $\beta$}}_a \delta N^a - \mbox{{\boldmath
$\lambda$}}^{ab} \delta \sigma_{ab}\right)\nonumber\\ &=& \int_{C_t}
t\cdot\left(N \delta\mbox{{\boldmath $\alpha$}} +
N^a\delta\mbox{{\boldmath $\beta$}}_a - \mbox{{\boldmath
$\lambda$}}^{ab} \delta \sigma_{ab}\right) \end{eqnarray} where eqs.
(\ref{H}), (\ref{thingy2}),
 (\ref{thdB}), and (\ref{Hexp})were used. Thus, we obtain
\begin{eqnarray} \delta I_m & = & \int_M {\bf E}^{ab} \delta g_{ab}-
\int_{\partial M} dt \wedge t\cdot\left( N \delta\mbox{{\boldmath
$\alpha$}} + N^a\delta\mbox{{\boldmath $\beta$}}_a - \mbox{{\boldmath
$\lambda$}}^{ab} \delta \sigma_{ab} \right) \nonumber\\ & = &\int_M
{\bf E}^{ab} \delta g_{ab} - \int_{\partial M} N
\delta\mbox{{\boldmath $\alpha$}} + N^a\delta\mbox{{\boldmath
$\beta$}}_a - \mbox{{\boldmath $\lambda$}}^{ab} \delta \sigma_{ab}
\end{eqnarray} which agrees precisely with eq. (\ref{deIMBY}).

Motivated by path integral methods for computing entropy, Brown and
York \cite{BY2} proposed a prescription for obtaining the entropy of
a black hole in general relativity. When generalized to the class of
theories we consider here, their prescription can be reformulated as
follows: Consider a Lorentzian black hole solution $(M, g_{ab})$ with
bifurcation surface ${\cal H}$, so that the spacetime manifold $M$
has topology ${\cal R}^2 \times {\cal H}$. Normalize the Killing
field $t^a$ which vanishes on ${\cal H}$ so that it has unit surface
gravity. Choose a slicing, $\Sigma_t$, of the exterior region of the
black hole labeled by Killing parameter $t$, such that each
$\Sigma_t$ smoothly intersects ${\cal H}$.  Define a ``Euclidean
manifold" $M'$ by writing $T = it$, taking $T$ to be real, and then
periodically identifying $T$ with period $2 \pi$ to avoid a conical
singularity at ${\cal H}$. We choose the boundary of $M'$ to be an
orbit of the Killing field, so that $\partial M'$ has topology $S^1
\times {\cal H}$.  Define a (in general, complex) ``Euclidean metric"
$g^E$ on $M'$ by analytic continuation of $g$.  Let $I'_m$ denote the
``Euclidean microcanonical action", defined by \begin{equation} I'_m
= -i \left(\int_{M'}{\bf L} - \int_{\partial M'} dt\wedge{\bf
Q}[t]\right) \label{EIm} \end{equation} where ${\bf L}$ and ${\bf Q}$
are analytically continued from the Lorentzian spacetime. (The
imaginary factor makes $I_m$ real since the ``$dt$" implicitly
appearing in ${\bf L}$ is imaginary on $M'$.) Then, the prescription
of Brown and York corresponds to the formula \begin{equation} S =
-I'_m \label{defSM} \end{equation}

We now show that -- in its domain of applicability -- eq.
(\ref{defSM}) is equivalent to the Noether charge prescription,
eq.(\ref{S}). Writing $T^a = - i t^a$ (so that $T^a \nabla_a T = 1$),
we have \begin{eqnarray} I'_m &=& -i \int_{M'}{\bf L} + i
\int_{\partial M'} dt\wedge{\bf Q}[t] \nonumber\\ & = &  -i
\int_{M'}dT \wedge T \cdot {\bf L} + \int_{\partial M'}dT \wedge {\bf
Q}[t] \nonumber\\ & = & - 2 \pi i \int_{\Sigma_0} T \cdot {\bf L}
-2\pi \int_{C_0} {\bf Q}[t] \nonumber\\ & = & - 2 \pi
(\int_{\Sigma_0} t \cdot {\bf L} + \int_{C_0} {\bf Q}[t])
\label{thingy4} \end{eqnarray} where the apparent sign change in the
surface term in the third line is due to our orientation conventions
(see the first paragraph of this section), and it should be noted
that the slice $\Sigma_0$ and cut $C_0$ are common to both $M$ and
$M'$.  However, we have \begin{eqnarray} \int_{\Sigma_0} t\cdot{\bf
L} &=&\int_{\Sigma_0}({\bf \Theta}(\phi,{\cal L}_t\phi)-{\bf J}[t])
\nonumber\\ &=&-\int_{\Sigma_0}d{\bf Q}[t] \nonumber\\ &=& -
\int_{C_0}{\bf Q}[t] + \int_{{\cal H}}{\bf Q}[t]  \nonumber\\
\label{LJQ} \end{eqnarray} Thus, we obtain \begin{eqnarray} S & = &
2 \pi \left(-\int_{C_0} {\bf Q}[t] + \int_{{\cal H}}{\bf Q}[t]  +
\int_{C_0}{\bf Q}[t]\right)\nonumber\\ & = &  2 \pi  \int_{{\cal
H}}{\bf Q}[t] \end{eqnarray} in agreement with eq.(\ref{S}).

\section{Black Hole Entropy as a Hilbert Action Boundary Term}
Ba\~nados, Teitelboim and Zanelli \cite{T1} have given another
approach for computing the entropy of black holes applicable to
general relativity and, more generally, Lovelock gravity. The
starting point of this approach is essentially the same as the
microcanonical action approach discussed in the previous section.
However, these authors then quote, without detailed derivation, the
following formula for the entropy of a stationary, Euclidean black
hole in Lovelock gravity \begin{equation} S = - \lim_{\epsilon
\rightarrow 0} \int_{\partial D_{\epsilon}\times {\cal H}}{\bf B}
\label{IT1} \end{equation} Here ${\bf B}$ is the Hilbert action
boundary form, defined by eq.(\ref{defB}) for variations which keep
the induced metric fixed on the boundary, and $D_{\epsilon}$ is a
two-dimensional disk of radius $\epsilon$ orthogonal to the
bifurcation surface ${\cal H}$ (on which the stationary Killing field
vanishes).

Our aim here is to explain why (\ref{IT1}) -- which looks very
different from either (\ref{defSM}) or (\ref{S}) -- actually produces
answers which coincide with these other calculations. To see this, we
consider a theory which satisfies the three assumptions of section
(2).  Let $\Sigma$ be a smooth hypersurface (transverse to the
Killing field $t^a$) in the Euclidean space which passes through the
bifurcation surface, and let ${\cal H}_\epsilon$ be a smooth,
one-parameter family of surfaces in $\Sigma$ which approach the
bifurcation surface ${\cal H}$ as $\epsilon \rightarrow 0$. Then, by
assumption (3) of section(2), ${\bf e}_a$ will be well defined on
each ${\cal H}_\epsilon$ and will smoothly approach its value on the
bifurcation surface as $\epsilon \rightarrow 0$.  However, it then
follows immediately from eq.(\ref{Hint}) that on the bifurcation
surface (where $t^a$ vanishes) we have $H_X = 0$. Thus, we find that
\begin{equation} \lim_{\epsilon \rightarrow 0} H_X (\epsilon) = 0
\label{limH} \end{equation}
{}From the original definition of $H_X$, eq.(\ref{H}), we thus obtain
\begin{equation} \lim_{\epsilon \rightarrow 0}\int_{{{\cal
H}}_\epsilon} {\bf Q}[t] = \lim_{\epsilon \rightarrow 0}\int_{{{\cal
H}}_\epsilon} t\cdot{\bf B} \end{equation} where the boundary surface
used to define ${\bf B}$ at each $\epsilon$ is taken to be the orbit
of $H_\epsilon$ under the action of $t^a$, i.e., it is simply
$\partial D_{\epsilon}\times {\cal H}$.  By eq.(\ref{S}) above, the
left  side is simply $S/2 \pi$.  Therefore we obtain \begin{eqnarray}
S &=& 2 \pi \lim_{\epsilon \rightarrow 0}\int_{{{\cal H}}_\epsilon}
t\cdot{\bf B}\nonumber\\ & = &  -\lim_{\epsilon \rightarrow 0}
\int_{\partial D_{\epsilon}\times {\cal H}} dt \wedge t\cdot{\bf
B}\nonumber\\ & = &  -\lim_{\epsilon \rightarrow 0} \int_{\partial
D_{\epsilon}\times {\cal H}} {\bf B} \label{thingy3} \end{eqnarray}
where the sign change in the second line results from our orientation
conventions as explained at the end of the first paragraph of section
3.  This establishes the equivalence of eq. (\ref{IT1}) and
(\ref{S}).

The explicit form of eq.(\ref{IT1}) for Lovelock gravity given in
\cite{T1} appears to be closely related to eq(\ref{ER}) above.  (That
those two formulae must be equivalent follows from eq.
(\ref{thingy3}) together with the equivalence of eqs.(\ref{S}) and
(\ref{ER}) proven in \cite{IW}.) If the Lagrangian does not contain
derivatives of the Riemann tensor, then ${\bf \Theta}$ can always be
chosen to have the form \cite{IW} \begin{equation} {\bf \Theta} =  2
\mbox{{\boldmath $\epsilon$}}_{aa_1...a_{n-1}} E_R^{abcd} {\bf
\nabla}_c \delta g_{bd} +{{\bf S}_{a_1...a_{n-1}}}^{bd}\delta g_{bd}
\label{Lth} \end{equation} where $E_R^{abcd}$ was defined below
eq.(\ref{ER}). Hence if a Hilbert action surface term, ${\bf B}$,
exists, one would expect there to be a direct relationship between it
and $E_R^{abcd}$. Thus it may be possible to give a simple direct
proof of the equivalence of eqs.(\ref{IT1}) and (\ref{ER}), although
we have not succeeded in doing so.

\section{Black Hole Entropy as a Quantity Conjugate to Conical
Deficit Angle} Ba\~nados, Teitelboim and Zanelli \cite{T1} and
Susskind and Uglum \cite{SU} have proposed another approach to
obtaining the entropy of a static black hole. The starting point of
this approach is the fact that in ordinary quantum field theory, the
partition function, $Z$, at inverse temperature, $\beta$, is given by
a path integral over Euclidean configurations with period $\beta$ of
$e^{- I}$, where $I$ is the Euclidean action. Thus, if we apply this
formula to a static black hole solution, then in the ``zero loop"
approximation, we simply have $Z = e^{- I}$, where $I$ is the usual
``Hilbert action" of the Euclidean black hole given by
\begin{equation} I_m = -i \int_{M'}{\bf L} + i\int_{\partial M'} {\bf
B} \end{equation} (see eq.(\ref{EIm}) above.) Thus, under these
assumptions, the Helmholz free energy, $F$, of a static black hole
would be given by \begin{equation} F = - \log Z/\beta = I/\beta.
\end{equation} The entropy of the black hole should then be given by
\begin{equation} S = \beta^2 \frac{\partial F}{\partial \beta} =
\beta \frac{\partial I}{\partial \beta} - I \label{SF} \end{equation}
where, in ordinary thermodynamics, the partial derivatives would be
taken at fixed values of the state parameters (other than energy). A
possible interpretation of the partial derivative in our case would
be to restrict consideration to variations in which the geometry is
changed only by varying the periodicity of the Euclidean time
coordinate. Since the contributions to $I$ from both the ordinary
volume term and the surface term at infinity are linearly
proportional to the periodicity, $\beta$, the variations of these
terms will not contribute to $S$ in eq.(\ref{SF}). However, when
$\beta$ is varied away from $\beta = 2 \pi$, a conical singularity is
created at the bifurcation surface of the black hole.  This conical
singularity can be viewed as corresponding to a $\delta$-function in
the curvature, which is proportional to $2 \pi - \beta$ rather than
$\beta$. Hence, this $\delta$-function contribution to the volume
term in $I$ will yield a nonzero contribution to $S$ in
eq.(\ref{SF}).

For the case of general relativity, the Lagrangian density is linear
in the curvature, so there is no difficulty in defining the
contribution of the $\delta$-function to the volume term in $I$.
Equation (\ref{SF}) then yields the standard result $S = A/4$
\cite{T1,SU}. However, if the Lagrangian density is a nonlinear
function of the curvature, it is far from clear that any well defined
regularization scheme can be given to define $I$ when a conical
singularity is present -- except in the limit in which the curvature
of the Euclidean black hole vanishes \cite{SU}, where the nonlinear
curvature terms of the Lagrangian can be neglected in any case.
Thus, it would appear that the conical deficit approach to
calculating the entropy of a black hole is fundamentally limited to
theories where the Lagrangian density is at most linear in the
curvature.

The relationship between the conical deficit and Noether charge
approaches to the calculation of black hole entropy has been analyzed
by Nelson \cite{Nelson}.  Making use of the fact that the variation
of the geometry resulting from a change of $\beta$ can be induced by
the action of a (singular) diffeomorphism, Nelson argues that in the
cases where the conical deficit approach can be defined, it should
yield the same result as the Noether charge method.

\section{Black Hole Entropy as Adduced From Pair Creation Rates} The
final Euclidean method for calculating black hole entropy which we
shall analyze involves the calculation of pair creation rates of
black holes by instanton methods \cite{GGS}. The idea here is to
compare the rate for pair creation of black holes with a
corresponding rate for the pair creation for objects which have the
same exterior field as the black hole but do not have a horizon. The
enhancement factor of the black hole rate should measure the number
of ``internal states" of the black hole, and, thus, determine its
entropy. The particular calculation done in \cite{GGS} involved the
comparison of the pair creation rates of Reissner-Nordstrom black
holes and monopoles in a magnetic field, but we now shall show that
this approach always yields results in agreement with the Noether
charge approach (see also \cite{B}).

The calculation of black hole entropy by the pair creation method
proceeds as follows: One first finds a Euclidean instanton solution
corresponding to the black hole pair creation process, and asserts
that the pair creation rate, $\Gamma$, is proportional to $e^{-I}$,
where $I$ is the Euclidean action of the instanton. One then finds an
instanton solution describing the monopole (or other object) pair
creation process, and asserts that the pair creation rate,
$\tilde{\Gamma}$, for these objects is proportional to
$e^{-\tilde{I}}$, where $\tilde{I}$ is the Euclidean action of this
instanton. The entropy, $S$, of the black hole is then given by
\begin{equation} S = \ln \frac{\Gamma}{\tilde{\Gamma}} = \tilde{I} -
I \label{SPC} \end{equation}

As usual, the instanton action will have a volume term and a surface
term from infinity. Since, by assumption, the instanton solutions
agree near infinity, the surface term contributions from $I$ and
$\tilde{I}$ will cancel in eq.(\ref{SPC}), so we need only consider
the volume term, which is of the form \begin{equation} I_{V} = -i\int
{\bf L} \end{equation} where ${\bf L}$ is an analytic continuation of
the Lorentzian Lagrangian $n$-form to Riemannian metrics (see
eq.(\ref{EIm}) above.) This volume term for both $I$ and $\tilde{I}$
can be evaluated as follows. Let $\Sigma$ be a hypersurface which
intersects each (circular) orbit of the timelike Killing field once
and only once. In the black hole case, $\Sigma$ will terminate at the
bifurcation surface of the black hole, so $\partial \Sigma$ will
consist of the bifurcation surface together with a two-surface at
infinity. In the monopole (or other object) case $\partial \Sigma$
will be comprised by only the two-surface at infinity. In either
case, we have by a calculation which parallels eqs.(\ref{thingy4})
and (\ref{LJQ}) above, \begin{eqnarray} I_{V} = -i \int_{M'} {\bf L}
& = & -i \int_{M'} dT \wedge T \cdot {\bf L}\nonumber\\
 & = & -2 \pi i \int_\Sigma T \cdot {\bf L}\nonumber\\ & = & -2 \pi
 \int_\Sigma t \cdot {\bf L}\nonumber\\ & = & -2 \pi \int_\Sigma
[{\bf \Theta} - {\bf J}]\nonumber\\ & = &  2 \pi \int_\Sigma d {\bf
Q}\nonumber\\ & = &  2 \pi \int_{\partial \Sigma} {\bf Q}
\end{eqnarray} where we used the fact that the instanton is
stationary (so that ${\bf \Theta}(\phi, {\cal L}_t \phi) = 0$) and is
a solution (so that ${\bf J} = d {\bf Q}$). Thus, for the black hole
instanton, we obtain \begin{equation} I_{V} =  2 \pi \left(
\int_\infty {\bf Q} - \int_{{\cal H}} {\bf Q} \right) \end{equation}
whereas for the monopole (or other object) we have \begin{equation}
\tilde{I}_{V} = 2 \pi  \int_\infty \tilde{{\bf Q}} \end{equation}
Since the instanton solutions agree at infinity, we have ${\bf Q} =
\tilde{{\bf Q}}$ there. Thus, eq.(\ref{SPC}) yields \begin{equation}
S = 2 \pi \int_{{\cal H}} {\bf Q} \end{equation} in agreement with
the Noether charge method, as we desired to show.

\section*{Acknowledgements} This research was supported in part by
National Science Foundation grant PHY-9220644 to the University of
Chicago.

\section*{Appendix: The Structure of the ``Off-Shell" Hamiltonian} In
this Appendix, we derive the general form of the Hamiltonian in a
diffeomorphism invariant theory of gravity.  Consider a theory
derived from a Lagrangian ${\bf L}$ on a manifold $M$, as described
above eq. (\ref{delL}). Suppose, in addition, that for the chosen
boundary conditions (which could be either suitable asymptotic
flatness conditions or conditions at a finite boundary, as considered
in section 2), there exist forms ${\bf B}$ and $\mbox{\boldmath
$\mu$}$ satisfying eqs. (\ref{defB}) and (\ref{mu}). Then the
Hamiltonian, $H$, conjugate to a time translation vector field $t^a$
is determined by \cite{IW} \begin{equation} \delta H =
\Omega(\phi,\delta\phi,{\cal L}_t\phi)= \int_{\Sigma_t} \delta {\bf
J}[t] -  \int_{C_t}t\cdot \delta{\bf B} \label{delH} \end{equation}
where $\phi$ is a solution to the equations of motion, but the
variation $\delta$ is to an arbitrary nearby configuration (in
contrast to the situation considered in eq.(\ref{omeg}) above, where
$\delta\phi$ was required to satisfy the linearised equations of
motion). By inspection, we have \begin{eqnarray} H & = &
\int_{\Sigma_t}  {\bf J}[t] - \int_{C_t} t \cdot{\bf B}\nonumber\\ &
= & \int_{\Sigma_t}  {\bf J}[t] - d{\bf Q}[t] + \int_{C_t} {\bf Q}[t]
- t \cdot{\bf B} \label{Hgen} \end{eqnarray} where ${\bf Q}$ is
determined (up to the ambiguities analyzed in \cite{IW}) by the
relation $d {\bf Q} = {\bf J}$ when $\phi$ satisfies the equations of
motion and, for the present, its definition is extended in an
arbitrary, local, covariant manner to $\phi$ which do not satisfy the
equations of motion.  Clearly (\ref{Hgen}) reduces to our previous
expression (\ref{H}) ``on shell", i.e., when $\delta \phi$ satisfies
the linearized equaitons of motion about $\phi$. We now shall show
that ${\bf Q}$ always can be defined ``off shell" so that the
integrand of the volume integral in eq. (\ref{Hgen}) takes the form
\begin{equation} {\bf J}[t]-d{\bf Q}[t] = t^a {\bf {\cal C}}_a
\label{JdmQ} \end{equation} where ${\bf {\cal C}}_a$ is locally
constructed out of the dynamical fields in a covariant manner and
${\bf {\cal C}}_a = 0$ when the equations of motion are satisfied.
Indeed, it follows from the analysis of \cite{LW} that we may view
${\bf {\cal C}}_a = 0$ as being the constraint equations of the
theory which are associated with its diffeomorphism invariance.
Thus, the general form of the Hamiltonian in a theory arising from a
diffeomorphism covariant lagrangian is \begin{equation} H =
\int_{\Sigma_t}  t^a {\bf {\cal C}}_a + \int_{C_t} {\bf Q}[t] - t
\cdot{\bf B} \label{Hlin} \end{equation}

To prove (\ref{JdmQ}), we note that the freedom in extending the
definition of ${\bf Q}$ off-shell clearly allows us to make the
replacement ${\bf Q} \rightarrow ({\bf Q} + \mbox{\boldmath
$\tau$})$, where $\mbox{\boldmath $\tau$}$ is any covariant $(n -
2)$-form locally constructed from the dynamical fields such that
$\mbox{\boldmath $\tau$} = 0$ whenever the equations of motion, ${\bf
E} = 0$, are satisfied.  To prove that this freedom is sufficient to
ensure that eq. (\ref{JdmQ}) holds, we recall \cite{IW} that ${\bf
J}[t] - d{\bf Q}[t]$ is a linear differential operator on the vector
field $t^a$, so we can write it as a sum \begin{equation} ({\bf
J}-d{\bf Q})_{a_1\ldots a_{n-1}}=\sum_{i=0}^{m} {}^{(i)}
{{A_{a}}^{b_1\ldots b_i}}_{a_1\ldots a_{n-1}} {\bf
\nabla}_{(b_1\ldots}{\bf \nabla}_{b_i)} t^a \label{JmdQ}
\end{equation} where the coefficients ${}^{(i)} {{A_{a}}^{b_1\ldots
b_i}}_{a_1\ldots a_{n-1}}$ are locally and covariantly constructed
from the dynamical fields and have the symmetries ${}^{(i)}
{{A_{a}}^{b_1\ldots b_i}}_{a_1\ldots a_{n-1}} = {}^{(i)}
{{A_{a}}^{(b_1\ldots b_i)}}_{a_1\ldots a_{n-1}} = {}^{(i)}
{{A_{a}}^{b_1\ldots b_i}}_{[a_1\ldots a_{n-1}]}$.  In fact, the
analysis of \cite{IW} shows that we may always choose ${\bf J}$ and
${\bf Q}$ so that $m = 2$, but this fact does not simplify the proof,
so we shall leave eq. (\ref{JmdQ}) in the form of a general sum.
Since ${\bf J}=d{\bf Q}$ when the field equations hold, we have
\begin{equation} {}^{(i)} {{A_{a}}^{b_1\ldots b_i}}_{a_1\ldots
a_{n-1}}=0 \mbox{ when } {\bf E} = 0.  \label{Aeom} \end{equation} We
now parallel the proof of lemma 1 of \cite{onclosed} to show that if
eq.(\ref{JmdQ}) holds with  $m \ge 1$, an $(n-2)$-form
$\mbox{\boldmath $\tau$}$ always can be chosen so that ${\bf J} - d
{\bf Q} - d \mbox{\boldmath $\tau$}$ is of the form of the right side
of eq.(\ref{JmdQ}), but with the sum terminating at $(m - 1)$. By
induction, it then will follow immediately that ${\bf Q}$ can be
chosen so that eq.(\ref{JdmQ}) holds.

To proceed, we recall that (see, e.g., \cite{LW}) \begin{equation} d
({\bf J}[t] - d {\bf Q}) = d{\bf J} = -{\bf E}{\cal L}_t\phi
\label{dJ} \end{equation} Now the right side of this equation
involves only one derivative of $t^a$. Consequently, substituting on
the left side from eq.(\ref{JmdQ}), and assuming $m\ge 1$ we obtain
\begin{eqnarray} 0 &=& {}^{(i)} {{A_{a}}^{(b_1\ldots
b_i}}_{[a_1\ldots a_{n-1}} \delta^{c)}_{d]} {\bf \nabla}_{(c} {\bf
\nabla}_{b_1\ldots}{\bf \nabla}_{b_m)} t^a \nonumber\\
&&+\mbox{(terms with fewer symmetrised derivatives of $t^a$)}
\label{dJmdQ} \end{eqnarray} Since this equation holds for all $t^a$,
it follows that \begin{equation} {}^{(i)} {{A_{a}}^{(b_1\ldots
b_i}}_{[a_1\ldots a_{n-1}} \delta^{c)}_{d]} = 0.  \label{dA}
\end{equation} By inspection, it then follows \cite{onclosed} that
\begin{equation} \mbox{\boldmath $\tau$}_{a_2\ldots a_{n-1}} =
\frac{m}{m+1} {}^{(i)} {{A_{a}}^{cb_2\ldots b_i}}_{ca_2\ldots
a_{n-1}} {\bf \nabla}_{b_2\ldots}{\bf \nabla}_{b_m}t^a \label{taudef}
\end{equation} satisfies the desired requirement that ${\bf J}-d{\bf
Q}-d\mbox{\boldmath $\tau$}$ has at most $m-1$ symmetrised
derivatives of $t^a$. Furthermore, eq. (\ref{Aeom}) implies that
$\mbox{\boldmath $\tau$} = 0$ whenever the equations of motion hold,
so the substitution ${\bf Q} \rightarrow ({\bf Q} + \mbox{\boldmath
$\tau$})$ does not affect the definition of ${\bf Q}$ ``on shell".


\begin{thebibliography}{99}
\bibitem{com1} Matter fields,
$\psi$, can be included straightforwardly if ${\cal L}_v \psi = 0$ on
$\Sigma_t$ for all $v^a$ which vanish on $\Sigma_t$.  For higher
derivative gravity theories, one undoubtedly would need to generalize
the choice of $X$ boundary conditions given in assumption (1) to
include the appropriate additional ``configuation" degrees of freedom
associated with the spacetime metric.
\end{thebibliography}
\end{document}